\documentclass[prl,aps,superscriptaddress]{revtex4}
\usepackage{graphicx}
\usepackage{epsfig}
\usepackage{amsmath}
\usepackage{amssymb}
\usepackage{tabularx}
\usepackage{multirow}
\usepackage{array}

\begin{document}    

\title{Modelling Dual Pathways for the Metazoan Spindle Assembly
Checkpoint}

\author{Richard P. Sear} 
\affiliation{Department of Physics, University of Surrey, Guildford,
Surrey GU2 7XH, UK.}

\author{Martin Howard} 
\affiliation{Department of Mathematics, Imperial College London, South
Kensington Campus, London SW7 2AZ, UK.}

\date{\today}
\begin{abstract}
Using computational modelling, we investigate mechanisms of signal
transduction 
focusing on 
the spindle assembly checkpoint where a single unattached
kinetochore is able to signal to prevent cell cycle progression. This
inhibitory signal switches off rapidly once spindle microtubules have
attached to all kinetochores. This requirement tightly constrains the
possible mechanisms. Here we investigate two possible mechanisms for
spindle checkpoint operation in metazoan cells, both supported by
recent experiments. The first involves the free diffusion and
sequestration of cell-cycle regulators. This mechanism is severely
constrained both by experimental fluorescence recovery data and also
by the large volumes involved in open mitosis in metazoan cells. Using
a simple mathematical analysis and computer simulation, we find that
this mechanism can generate the inhibition found in experiment but
likely requires
a two stage signal amplification cascade.
The second mechanism involves spatial gradients of a short-lived
inhibitory signal that propagates first by diffusion but then
primarily via active transport along spindle microtubules.  We propose
that both mechanisms may be operative in the metazoan spindle assembly
checkpoint, with either able to trigger anaphase onset even without
support from the other pathway.
\end{abstract}

\maketitle 


\vspace*{0.8cm}

The question of how a signal emanating from a small, compact structure
in a cell can be amplified and propagated to an entire cell is
fundamental to cell biology \cite{nasmyth05}. An excellent example is
provided by the spindle assembly checkpoint (SAC) \cite{musacchio02},
which regulates cell cycle progression from metaphase to anaphase
during mitosis. The segregation of sister chromatids that occurs
during anaphase is permitted only after all the kinetochores are
attached via microtubules to the mitotic spindle. Even a single
unattached kinetochore can signal to the rest of the cell and prevent
cell cycle progression \cite{rieder95,rieder94}. A fundamental issue
is how a relatively small structure, such as a kinetochore, can
generate sufficient signal to robustly communicate with distant
subcellular locations \cite{nasmyth05}. Moreover, this signal must
switch off rapidly, within a period of minutes, after complete
kinetochore attachment to spindle microtubules \cite{rieder94}. These
requirements strongly constrain the possible signal transduction
mechanisms.  In this paper, we focus
particularly on the SAC in cases where the nuclear envelope breaks
down prior to SAC activity (open mitosis), as in metazoan cells. In
this context, we examine two distinct models: a diffusive
sequestration model and a model involving active signal transport
along spindle microtubules.  We believe that both of these pathways
may be in simultaneous operation in the metazoan SAC. 

The metaphase/anaphase transition is triggered by an intricate
sequence of events centred around the proteins securin, cyclin B and
separase.  The first step is the ubiquitination of securin and cyclin
B by the Anaphase Promoting Complex/Cyclosome (APC/C) \cite{castro05},
a process that tags securin/cyclin B for destruction via the 26S
proteasome. This degradation allows separase to cleave the cohesin
complex that tethers sister chromatids together. Once the cohesin
complex has been cleaved, the sister chromatids aree pulled apart to
opposite poles by the mitotic spindle. In order to prevent premature
entry into anaphase, the SAC must prevent securin/cyclin B
ubiquitination by the APC/C until proper attachment of {\it all}
chromosomes to the spindle. Evidence has accumulated for a number of
overlapping, and therefore possibly redundant, mechanisms for SAC
operation. The APC/C is known to be stimulated by Cdc20 binding;
hence, a plausible way to achieve APC/C inhibition is to inhibit the
ability of Cdc20 to bind to the APC/C. One possibility is that Cdc20
is held, sequestered, in an inactive form via binding to Mad2, with
the production of Cdc20-Mad2 being promoted by unattached kinetochores
\cite{howell04,deantoni051,deantoni052}. Key proteins identified at
unattached kinetochores include Bub1, Mad1, Mad2, BubR1 (Mad3 in
budding yeast), Bub3 and Cdc20. Moreover, FRAP experiments have
revealed that some of these proteins, including Mad2, BubR1 and
Cdc20, turnover rapidly at unattached kinetochores
\cite{howell04,shah04}. Furthermore, the available evidence suggests
that Mad2 exists in two forms: open (O-Mad2) and closed (C-Mad2)
\cite{sironi02,luo02}, with the closed form adopted when bound to
Cdc20.  Production of C-Mad2-Cdc20 may be catalysed via the
kinetochore bound C-Mad2-Mad1 complex. Intriguingly,
Refs.~\cite{deantoni051,deantoni052} propose that C-Mad2-Cdc20 away
from the unattached kinetochore can convert further cytosolic O-Mad2
and Cdc20 into their bound C-Mad2-Cdc20 state. In this way the
relatively weak signal coming from an unattached kinetochore can be
amplified leading to comprehensive Cdc20 sequestration throughout the
cell. Recent experiments have further implicated a protein called
p31$^{\rm comet}$ in switching off this signal after complete
kinetochore attachment \cite{deantoni051,xia04,mapelli06}.

However, the above ``Mad template'' model is not the only proposed
mechanism for APC/C repression. BubR1 and Bub3 are also known to bind
Cdc20 and thus repress the APC/C \cite{tang01}. Indeed, BubR1 appears to
be a more potent inhibitor of Cdc20 than Mad2, and both BubR1 and Mad2
may mutually promote each other's binding to Cdc20 \cite{fang02}.
Furthermore, Bub1 is believed to phosphorylate Cdc20, possibly also
repressing the APC/C \cite{tang04}.  In addition, the overall copy
number of Cdc20 is down-regulated until all the kinetochores are
attached \cite{pan04}. Clearly, reducing the overall number of Cdc20
will impair the effectiveness of the APC/C prior to anaphase. Moreover,
not only does Cdc20 form complexes with Mad2 and BubR1, but it is also
believed to form a separate complex called the Mitotic Checkpoint
Complex (MCC), consisting of Mad2, BubR1, Bub3 and Cdc20
\cite{sudakin01}.  This complex also appears to be a potent inhibitor
of the APC/C \cite{sudakin01}. In addition, the APC/C is regulated by
phosphorylation through the kinase Cdk1 \cite{kraft03}.  Lastly, the
APC/C itself has intriguing localization properties, localizing, for
example, to unattached kinetochores \cite{acquaviva04}. This
positioning may also have implications for the mechanism of APC/C
inhibition.

Recently, a pioneering paper by Doncic {\it et al.}~\cite{doncic05}
introduced a careful mathematical analysis to the question of how the
SAC functions. Focusing on yeast, they analysed various simplified
models of how an unattached kinetochore can signal to the rest of the
cell. Yeast undergoes closed mitosis, where even at the onset of
anaphase the nuclear envelope is still complete. As a consequence, a
diffusive signal from an unattached kinetochore is only required to
propagate within the nuclear volume of a few cubic microns.  However,
one important difference between yeast and metazoan cells is that the
latter
undergo open mitosis, where the nuclear envelope has
broken down by the time the SAC is active; hence any freely diffusing
signal from an unattached kinetochore must propagate throughout the
cytoplasm, a volume a thousand times larger than the nucleus of a
yeast cell. 
Furthermore a considerable amount of data is available concerning the
metazoan SAC in the form of flourescent bleaching data and estimates
for the copy numbers of the relevant molecules.  In this paper, we
therefore use computational modelling to analyze the SAC in metazoan
cells.  We develop a new model that incorporates two step signal
amplification, a mechanism with close similarities to previously
investigated multistep signal cascades
\cite{heinrich02,kholodenko06}. This similarity points towards a close
connection between the SAC and other cell signalling systems, such as
MAPK cascades. Importantly, we find that our model generates robust
inhibition in metazoan cells, unlike models without an amplification
step.

As discussed above, the most popular models of APC/C inhibition
involve transport solely by diffusion.  However, these models are hard
to reconcile with experiments of Rieder {\it et
al.}~\cite{rieder97}. Using cells with two independent spindles,
Rieder {\it et al.}~found that unattached kinetochores on one spindle
did not block anaphase onset in a neighbouring complete spindle. This
experiment appears to indicate that any ``wait anaphase'' signal would
have to be limited to a single spindle, and {\it not} diffuse
throughout the cytosol. This finding is in clear contradiction with
the mechanisms introduced above.  For this reason we develop a second
model where a short-lived signal from an unattached kinetochore first
diffuses to a nearby microtubule, before being actively transported
along spindle microtubules. In this way the signal is restricted to
the spindle.  Moreover, we propose that both the active transport and
diffusive mechanisms may be in simultaneous use in metazoan cells. We
suggest that either of the signalling mechanisms may individually be
able to initiate anaphase, even without support from the other
pathway. With this assumption, our models are then fully consistent
with both the experiments of Rieder {\it et al.}~\cite{rieder97} and
with a diffusive sequestration model \cite{deantoni051}.

\section*{Results}

\section*{Mechanisms with diffusive transport}

We analyze
the case where the kinetochores control the concentration of a freely
diffusing species $c$. When a kinetochore is unattached and so is
signalling, a large majority of the species $c$ is in the $c^*$ state,
but when the last kinetochore itself switches off, the $c^*$ species
rapidly decays to the $c$ state, thus communicating the switch-off
(attachment) to the rest of the cell.  
We will sometimes refer to the $c^*$ species as being in the
inhibiting state as it is this state that prevents securin/cyclin B
ubiquitination by the APC/C. 

\

\noindent
{\bf The concentrations of the freely diffusing species are almost
uniform} The timescale for diffusion across the cell is
$\tau_D=r_c^2/D$, whereas the mean time for a molecule to collide with
the kinetochore is $\tau_C=r_c^3/(Dr_k)$. Here $D$ is the diffusion
constant for the protein; $r_c$ is the distance across the cell
(typically a few tens of microns); and $r_k\simeq 0.2\mu$m
\cite{mcewen98} is the radius of the kinetochore.
If we assume that the lifetime for the inhibiting $c^*$ species is
much longer than $\tau_D$ then any gradient in its concentration will
clearly be small. Furthermore, the timescale for diffusion across the
cell is much smaller than that for collisions with the kinetochore,
$\tau_D/\tau_C=r_k/r_c \ll 1$.  Hence, each molecule of the $c$
species criss-crosses the cell many times between kinetochore
reactions and so gradients in its concentration are also small in most
of the cell. If, on the other hand, the lifetime of the inhibiting
species is short with respect to $\tau_D$, then its concentration will
no longer be uniform and instead density gradients will form. This
scenario will arise later on when we consider models with active
transport.
However, for longer lifetimes, we can model the inhibition using
simple ordinary differential equations. 

\

\noindent
{\bf Inhibitor production only at a kinetochore} In the simplest
possible model, the inhibiting $c^*$ species is produced only at
unattached kinetochores.  We denote the steady state rate of
production of $c^*$ at the final unattached kinetochore by $J_{\rm
off}$.  In order to allow the inhibition to be switched off at the
beginning of anaphase, the inhibiting $c^*$ species must be
unstable. We model this instability by a first-order decay $c^*\to c$,
with rate constant $\alpha$.  Cdc20 is known to activate the APC/C and
so trigger anaphase.  Since both Mad2 and BubR1 are known to bind to
Cdc20, we can tentatively identify $c$ as Cdc20 while $c^*$ is a
Mad2-Cdc20 or BubR1-Cdc20 complex. As the kinetochore is such a small
structure, we now address the question of whether the flux of $c^*$
produced {\it only} at a single unattached kinetochore is sufficient
to maintain inhibition.
At steady state the fluxes on and off a kinetochore must be the same,
i.e. $J_{\rm on}=J_{\rm off}$. Furthermore, at steady state, the rate
of production of $c^*$ at the last unattached kinetochore must equal
its rate of decay, i.e. $J_{\rm off}=\alpha N_c^*$, where $N_c^*$ is
the number of inhibiting $c^*$ molecules.  Experimental data, largely
on marsupial PtK$_1$ and PtK$_2$ cells, gives estimates for many of
the model parameters \cite{howell00,howell04}. From this data we will
be able to estimate the value of $N_c^*$ and see whether good
inhibition can be obtained.

Experimental evidence for the flux $J_{\rm off}$ is available from
FRAP (Fluorescence Recovery After Photobleaching) data for the
recovery of fluorescence of checkpoint proteins at unattached
kinetochores. This data provides a direct measure of dissociation
rates. Together with an estimate of the copy number of each
kinetochore bound protein, we can then derive the off flux $J_{\rm
off}$.  For BubR1 the (fast phase, see \cite{howell04}) FRAP half-life
($t_{1/2}$) is 3s, whereas for Mad2 the half-life is $10-20$s
\cite{howell04,shah04}, with a kinetochore occupancy $N_k$ of about
1000 molecules for both \cite{howell04}. This gives an off flux of
about $J_{\rm off}=N_k \ln 2/t_{1/2} \sim 1000\times\ln 2/3\sim
200~s^{-1}$ for BubR1 and $J_{\rm off}\sim 50~s^{-1}$ for Mad2.
Note that these fluxes are actually the maximum possible, as not all
of the released molecules will be in the inhibiting form. 
The lifetime of the inhibiting species $\alpha^{-1}$ has not been
directly measured. However, anaphase begins about 20 min after the
last kinetochore becomes attached \cite{rieder94}. Clearly, the number
of inhibiting $c^*$ molecules must decrease dramatically over a period
of no more than about 10 min, giving an upper bound of
$\alpha^{-1}\sim 10$min, a value which is also consistent with the
data of Ref.~\cite{clute99}. If we sum the fluxes of Mad2 and BubR1
and set $\alpha^{-1}=10$ min, the number of molecules $N_{c*}=J_{\rm
off}\alpha^{-1}\simeq 150,000$.  The total number of Cdc20 molecules
is approximately $800,000$ \cite{howell04}. Thus with the experimental
fluxes and even assuming a decay rate as slow as is reasonable we find
that at steady state a single kinetochore can only sequester
approximately 20\% of the Cdc20.
Furthermore, if the lifetime of the inhibiting $c^*$ molecules is
shorter than supposed by our upper bound $\alpha^{-1}\sim 10$min, or
if the flux $J_{\rm off}$ is lower, then an even smaller fraction will
be in the inhibiting state.
Hence, the formation of Mad2/Cdc20 and BubR1/Cdc20 complexes solely at
an unattached kinetochore cannot maintain good inhibition.

The experimental FRAP data used in the above analysis was not
specifically derived from the {\it final} unattached
kinetochore. Potentially the final unattached kinetochore might have a
higher turnover rate, thereby generating greater inhibition. We
therefore develop a complementary approach to determine whether a
reaction only at a single unattached kinetochore is {\it in principle}
sufficient to sequester most of the Cdc20. We estimate the inhibition
that would be generated if the on flux $J_{\rm on}$ at the kinetochore
were maximised, i.e., diffusion limited with a diffusion constant $D$
at the top of the range of the values measured {\it in vivo}.
Diffusion constants in the cytoplasm of metazoan cells have typically
yielded values in the range $1$ to $30\mu$m$^2$s$^{-1}$
\cite{wojcieszyn81,lang86,seksek97,schmiedeberg04}. Also, Howell {\it
et al.} \cite{howell04} found that a bleached spot of cytoplasmic
GFP-Cdc20 $0.8\mu$m across recovered its fluorescence in less than
$0.2$s, implying a diffusion constant $\gtrsim 5\mu$m$^2$s$^{-1}$.  If
the on flux to the unattached kinetochore is diffusion limited, then
using a plate geometry appropriate for a kinetochore, the on flux at
steady state is given by $J_{\rm on} \sim 4 D r_k N_c/v_c$
\cite{berg93}. Here, $N_c$ is the number of $c$ molecules,
$v_c=6000\mu$m$^3$ is the cytoplasmic volume \cite{howell04}, and we
use a fast diffusion constant $D=30\mu$m$^2$s$^{-1}$. At steady state
the flux $J_{\rm on}$ must be balanced by the $c^*\to c$ flux,
$J_{c*\to c}=\alpha N_{c*}$. Equating these two fluxes and defining
the constant total number of molecules as $N=N_c+N_{c*}$ (neglecting a
negligible number at the kinetochore), we have $N_{c*}/N=(1+\alpha
v_c/4Dr_k)^{-1}\approx 3/4$. Hence, even in an optimistic scenario,
only about three-quarters of the Cdc20 can be
sequestered. Furthermore, to reach this limit, high flux rates are
required, much higher than found experimentally for BubR1/Mad2 (see
above).  For these reasons, it is unlikely (though still theoretically
possible) that the metazoan SAC could function through this method.

The fundamental difficulty with this mechanism can be presented in an
alternative way that brings out the essential role played by the cell
volume. The time each molecule spends in the $c^*$ form is on the
order of $\alpha^{-1}$. The time taken for each molecule in the
$c$ state to find the unattached kinetochore is about
$v_c/k_k$, where $k_k$ is the unattached kinetochore binding
rate. This time scale is proportional to the cell volume and hence
becomes longer as the cell size increases. As the volume increases,
for fixed lifetime $\alpha^{-1}$, each molecule spends less and less
time in the inhibiting state and eventually the mechanism
fails. 

\

\noindent
{\bf Autocatalysis} One way to try to increase the number of inhibiting
molecules would be for the reaction that sequesters Cdc20 to occur not
just on the kinetochore but also off the kinetochore.  If $c^*$ itself
catalyzes an off kinetochore $c\to c^*$ reaction then the reaction is
autocatalytic. Autocatalysis was considered by Doncic {\it et al.}
\cite{doncic05}, where it was found to be unsatisfactory for the case
of closed mitosis in yeast. The reason is that, for good inhibition,
the off-kinetochore autocatalysis will likely have a large reaction
rate, as the on-kinetochore reaction is on its own insufficient to
maintain good inhibition (see above). Consequently, when the
kinetochore reaction is switched off, the result is only a weak
perturbation of an autocatalytic reaction. In other words
inhibition of a ``go
anaphase'' signal either cannot be switched off, or switches off only
very slowly (see Fig.~1). 
Interestingly, the Mad2 template model
proposed by Ref.~\cite{deantoni051} is essentially identical to this
mechanism, where the closed form of Mad2 catalyzes the
conversion of open to closed Mad2 both on and off the
kinetochore. Thus an unattached kinetochore inhibits anaphase by
generating C-Mad2-Cdc20 which in turn generates more C-Mad2-Cdc20 off
the kinetochore via an autocatalytic reaction. However, by itself this
autocatalytic mechanism does not allow switch off of the metazoan SAC
for the same reason it does not for the SAC in yeast.

Autocatalysis could play a role in the inhibition, provided an
additional process switches off the autocatalytic reaction once the
last kinetochore attaches to a microtubule.  It has been suggested
that the protein p31$^{\rm comet}$ could play a role in this context
\cite{deantoni051,xia04,mapelli06}.  Potentially p31$^{\rm comet}$
could be upregulated to abruptly switch off the SAC. However, the
question is then how this signal could be turned on so rapidly after
microtubule attachment to the final kinetochore.
If p31$^{\rm comet}$ competes with O-Mad2 for C-Mad2 binding
\cite{mapelli06}, then this is no easier than regulating the
concentration of free Cdc20, our original problem.  Potentially,
p31$^{\rm comet}$ could be upregulated using a two step reaction
process; however such a scheme would have to be conceptually similar
to that discussed in the next section. We emphasize that p31$^{\rm
comet}$ could also play other important roles, for example in
switching off kinetochore signalling following microtubule attachment
\cite{vink06}.

\

\noindent
{\bf Model with an amplification step} 
We now turn to a
model with an off-kinetochore, but non-autocatalytic, reaction. This
involves the species: $e$, $e^*$ and $c^*$.  We first assume that an
$e$ species cycles through the unattached kinetochore, where it is
converted to the inhibiting $e^*$ form. We further assume that the
$e^*$ is able to convert the $e$ form to a second
inhibiting $c^*$ species, 
giving a second step to the signal amplification.  
Importantly the $c^*$ form cannot convert further $e$
into the $c^*$ or $e^*$ forms. This form of amplification ensures that
only molecules that have passed through the
unattached kinetochore participate in amplifying the inhibitory
signal. As the inhibitory signal is not produced autocatalytically, it
can switch off rapidly after the final kinetochore-microtubule
attachment. The reaction processes are:
\begin{eqnarray}
e+K&\stackrel{k_k}{\rightarrow}&e^*+K  \qquad
e^*~\stackrel{\alpha_e}{\rightarrow}~e \nonumber \\
e+e^*&\stackrel{k}{\rightarrow}&c^*+e^* \qquad
c^*~\stackrel{\alpha}{\rightarrow}~e, \label{twospec}
\end{eqnarray}
where $K$ is an unattached kinetochore.  It is natural to associate
the $e$ species with free Cdc20, and the $e^*$ and $c^*$ species with
complexes of Cdc20 such as C-Mad2-Cdc20 and BubR1-Cdc20.  We are
assuming that one form of the complex ($e^*$) can catalyze the
production of further complexes ($c^*$) which differ in that the $c^*$
cannot participate in manufacturing further complexes. This
distinction between the two forms is a clear prediction of our
modelling. 
However, we cannot definitively identify the difference between
the $c^*$ and $e^*$ species, which could, for example, involve
phosphorylation or a conformational change.  Clearly, our reaction
scheme is schematic; more complex schemes based on the same principles
are certainly feasible. 
One possibility is that there could be more than two steps to the
amplification process. Alternatively,
p31$^{\rm comet}$ could be
rapidly upregulated using a two step reaction, as discussed
previously. Another possibility is using a two step reaction to
modulate the decay rate $\alpha$. Both possibilities could in
principle lead to both good inhibition and rapid checkpoint switch
off, but are fundamentally similar to the scheme of
Eq.~{\bf\ref{twospec}}.  The key prediction of the model
of Eq.~{\bf\ref{twospec}} is
for at least two species allowing non-autocatalytic amplification off
the kinetochore and hence strong sequestration.

The above two-step process has close similarities to other multi-step
signalling cascades such as for MAPK
\cite{heinrich02,kholodenko06}. However, there are differences, for
example, the $e$ species participates in both steps of the above
amplification process.  In MAPK cascades, on the other hand, a
separate $c$ species is converted to the $c^*$ form in the second
amplification reaction.  Nevertheless, the principle of using more
than one step to provide robust amplification but with rapid response
times is similar and is likely conserved across many different
signalling systems. However, the difficulty of robust signalling in
the SAC is particularly acute, since the initial signal emerges from
such a small region (a single unattached kinetochore).

For the parameter values of the reactions in Eq.~{\bf\ref{twospec}},
we use $\alpha^{-1}=\alpha_e^{-1}=5 {\rm min}$. These lifetimes are
shorter than used previously: the two step nature of the reaction
mechanism now dictates that shorter lifetimes are needed for switch
off within the appropriate timeframe of less than $20$ min.  Despite
these short lifetimes, a two step reaction cascade ensures that robust
signal amplification is still achieved. In dilute solution {\it in
vitro}, rate constants for diffusion-limited protein-protein
association are around $10^{-3}$ to $10^{-2}\mu$m$^3$s$^{-1}$
\cite{northrup92}, with the exception of some large rate constants
where there is significant electrostatic attraction between the
proteins. We take an {\it in vivo} rate constant at the top end of
these values, with $k=10^{-2}\mu m^3 s^{-1}$. For the rate constant
$k_k$, we use the diffusion limited value $k_k\sim 4 D r$
\cite{berg93}. Assuming $D=20 \mu$m$^2 s^{-1}$, i.e. fairly fast
cytoplasmic diffusion, we find $k_k\sim 20 \mu $m$^3 s^{-1}$.

\begin{figure}[t!]
\begin{center}
\epsfig{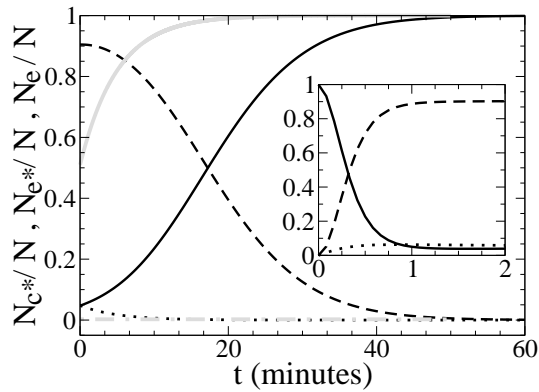}
\end{center}
\caption{
\lineskip 2pt \lineskiplimit 2pt Plot, for the model of
Eq.~{\bf\ref{twospec}}, showing the fraction of molecules $N_e/N$
(solid line), $N_{c*}/N$ (dashed line), $N_{e*}/N$ (dotted line) as a
function of time, with the final kinetochore attaching at time $t=0$,
i.e.~initial steady-state concentrations, but $k_k$ then set to zero
at $t=0$.  Inset shows the same fractions, but starting with all $N$
molecules in the $e$ form and with $k_k$ set to $20\mu$m$^3$s$^{-1}$
at $t=0$.  For comparison, in the main plot, the fractions $N_c/N$ for
the production only at kinetochores and autocatalytic models are
plotted as grey curves. They should be compared with the solid black
curve.
The grey solid curve is with a reaction only at a kinetochore,
and the grey dot-dashed curve is with autocatalysis. 
Note that with a
reaction only at a kinetochore the initial inhibition is weak whereas
with autocatalysis inhibition is near total even when $k_k$ is set to
zero.  
The values of $k$ (for the autocatalytic rate), $\alpha$ and
$k_k$ are the same in the single species models
as for the model of Eq.~{\bf\ref{twospec}}.
\label{dplot}
}
\end{figure}

Since the gradients in the concentrations are small (see above), we
can determine the time dependence and steady-state values of the three
species from ordinary differential equations:
\begin{eqnarray}
\frac{{\rm d}N_{e*}}{{\rm d}t}&=&\frac{k_k}{v_c}N_e-\alpha_e N_{e*},
\qquad \frac{{\rm d}N_{c*}}{{\rm d}t}=\frac{k}{v_c}N_{e*}N_e
-\alpha N_{c*},
\label{modeld2}
\\ 
\frac{{\rm d}N_{e}}{{\rm d}t}&=&-\frac{k_k}{v_c}N_e-\frac{k}{v_c}
N_{e*}N_e + \alpha_e N_{e*}+\alpha N_{c*},
\label{modeld3}
\end{eqnarray}
where $N_x$ is the number of molecules of species $x=e,e^*,c^*$. At
steady state (ss), we have:
\begin{equation}
N_{e*}^{ss}=\frac{k_k}{v_c\alpha_e}N_e^{ss}, \qquad
N_{c*}^{ss}=\frac{k k_k}{v_c^2 \alpha \alpha_e}\left(N_e^{ss}\right)^2. 
\end{equation}
Defining the (constant) total number of molecules as
$N=N_e+N_{e*}+N_{c*}$, we find:
\begin{equation}
N=\left(1+\frac{k_k}{v_c \alpha_e}\right)N_e^{ss}+
\frac{k k_k}{v_c^2\alpha \alpha_e}\left(N_e^{ss}\right)^2.
\end{equation}
For $N=800,000$, and using the above parameters, we find
$N_e^{ss}\approx 40,000$, $N_{e*}^{ss}\approx 40,000$ and
$N_{c*}^{ss}\approx 720,000$. Hence around $95$\% of the molecules are
in the inhibiting state.  For the single unattached kinetochore, we
find that the diffusion limited on rate onto the kinetochore is about
$J_{\rm on}\sim 4 D r N_e^{ss}/v_c\sim 100s^{-1}$. Assuming a
kinetochore population of around $1000$, as found experimentally, this
gives a half-life for the kinetochore bound population of
$5-10$s. This is roughly consistent with the observed Mad2, BubR1
kinetochore half-lives \cite{howell04,shah04}. As shown in
Fig.~\ref{dplot}, we also find that the signal switches off quite
quickly after microtubule attachment to the final kinetochore.  After
10 min the fraction of the $e$ molecules has increased from 5\% to
24\% and after 20 min 60\% is in the $e$ form. Moreover the
switch-{\it on} of the checkpoint is even quicker, with good
inhibition being reestablished within 1 min of even a single
kinetochore detachment, see Fig.~\ref{dplot} inset.  This is in good
agreement with cyclin B1 data from Ref.~\cite{clute99}, which suggests
that switch-on is essentially an order of magnitude faster than SAC
switch-off.

We therefore conclude that the above model is compatible with
experiments. However, we did use a diffusion-limited value for $k_k$
and a value of $k$ near the top end of the range of reaction-rate
constants for typical proteins in a dilute solution
\cite{northrup92}. Reducing $k$ by an order of magnitude weakens the
level of inhibition, although over $85\%$ of the molecules are still
in the inhibiting state. However, the model is not consistent with
experiments if the reaction rates are further reduced or if diffusion
is substantially slowed. Thus we predict that if the mechanism of the
SAC is diffusive amplified sequestration then measurements of the
reaction rates will reveal rather fast reactions,
or possibly more than two steps to the amplification process.

For completeness we also analyze an alternative two species model
previously proposed by Doncic et al \cite{doncic05}, for the smaller
volumes involved in the yeast SAC. The reaction scheme
for their model is:
\begin{eqnarray}
 e+K&\stackrel{k_k}{\rightarrow}&e^*+K  \qquad
 e^*~\stackrel{\alpha_e}{\rightarrow}~e \nonumber \\
 c+e^*&\stackrel{k}{\rightarrow}& 
 c^* \hspace{1.475cm} c^*~\stackrel{\alpha}{\rightarrow}~c+e.
\label{twospecdoncic}
\end{eqnarray}
Note that this scheme does not catalytically amplify the inhibitory
signal.  Here, one $e^*$ molecule can only interact with one $c$ molecule
while in our previous model a single $e^*$ molecule can convert many
molecules into the inhibiting form, thereby producing
amplification. For good inhibition we require
$N_{e^*}^{ss}+N_{c*}^{ss}\approx 800,000$. Using
$\alpha^{-1}=\alpha_e^{-1}=5{\rm min}$, we find that the flux off the
unattached kinetochore must be well over $1000s^{-1}$, faster than the
diffusion limited maximum, even for high levels ($10^5$ copies) of the
$e$ species. This finding is, of course, not unexpected: the
lack of amplification means that the flux of inhibitory molecules off
the unattached kinetochore must be higher than in the catalytic model
proposed above.  We therefore believe that this model is probably not
able to account for the SAC in metazoan cells.

\section*{Mechanisms involving active transport}

As shown above, our model with an amplification step is able to
explain many features of the metazoan SAC.  However, it is not
consistent with the experiments of Rieder {\it et al.}
\cite{rieder97}.  They observed that an incomplete spindle did {\em
not} inhibit another complete spindle 20$\mu$m away. 
Furthermore an unattached kinetochore was found to inhibit anaphase
onset everywhere within its local spindle even those parts more than
20$\mu$m away. These findings are clearly incompatible with models
where inhibition propagates purely diffusively away from incomplete
kinetochores into the cytoplasm.

These observations motivate us to consider mechanisms in which the
spindle itself plays an active role. If the signal is propagated
within the spindle itself, and not spread throughout the cytoplasm,
then the obvious transport mechanism is via molecular
motors. Propagation via active transport along microtubules is fast;
motors can move at speeds of microns per second \cite{howard01}. As
the typical spindle length scale is approximately 10$\mu$m, then
transport across the spindle takes only seconds at that speed,
consistent with rapid checkpoint switch-on/off. Of course, by
definition, an unattached kinetochore is not connected via
microtubules to a spindle pole. Hence that kinetochore must first
produce a freely diffusing species to carry the signal as before.
This inhibitory species, which we denote by $g^*$, initially diffuses
through the cytoplasm, but only until it either encounters a minus-end
directed microtubule-bound molecular motor, or decays to an inactive
form, $g$. If the molecule encounters a microtubule-bound motor, this
binding then stabilises the active $g^*$ form and transports the
inhibitory signal to a spindle pole.

However, before we can conclude that this model with active transport
is consistent with the experimental data of Rieder {\it et
al.}~\cite{rieder97}, we need to demonstrate that it is possible to
find a lifetime for the $g^*$ that is long enough to allow it to
encounter a motor but short enough to prevent more than a small
fraction diffusing 20$\mu$m away. If $g^*$ is manufactured at a rate
$J_{g*}$ at unattached kinetochores and decays at a rate
$\alpha_{g*}$, then the concentration $c_{g*}({\bf r},t)$ satisfies
the partial differential equation
\begin{equation}
\frac{\partial c_{g*}}{\partial t}=D\nabla^2 c_{g*}-
\alpha_{g*} c_{g*}+J_{g*}\delta^3(r).
\label{rd}
\end{equation}
We solve Eq.~{\bf\ref{rd}} at steady state, after assuming spherical
symmetry around the source (the kinetochore) and negligible
concentration of $g^*$ at large distances from the source. The
solution is
\begin{equation}
c_{g*}(r)=(J_{g*}/4\pi) \left(\lambda/r\right)\exp(-r/\lambda),
\end{equation}
where $\lambda=\sqrt{D/\alpha_{g*}}$ and $r$ is the distance from the
kinetochore. Of course, the assumption of spherical symmetry is a
gross approximation, especially as the kinetochore is a plate shaped
structure. However, we are only interested in qualitative results for
which this approximation will be reasonable at large distances. If we
set the cytoplasmic $g^*$ lifetime to be $\alpha_{g*}^{-1}=0.5$s, then
even with a large diffusion constant of $D=20\mu$m$^2$s$^{-1}$,
$\lambda\sim 3\mu$m. Hence the signal $20\mu$m away will be greatly
attenuated.
Due to the short lifetime of the $g^*$ form, the inhibitor forms a
steep gradient inside the cell. Hence we predict that subcellular
concentration gradients, already believed to be important for
microtubule growth and kinetochore capture, also play an important
role in checkpoint function \cite{wollman05}.

When we considered diffusive sequestration in the previous section we
found kinetochore fluxes on the order of 100 molecules/s.  If we
assume a similar flux $J_{g*}=100/$s then the concentration of $g^*$
$1\mu$m away from the source is approximately $20\mu$m$^{-3}$. At this
concentration the reaction rate per motor is $20k$, where $k$ is the
reaction rate between a pair of proteins. If, as in the previous
section, we take $k=10^{-2}\mu$m$^3$s$^{-1}$ \cite{northrup92}, then
we have a rate per motor complex of $0.2$s$^{-1}$. Thus a motor will
pick up a $g^*$ molecule within a few seconds if it is close to an
unattached kinetochore. If we assume a motor density of
$10\mu$m$^{-1}$, moving at $1\mu$m s$^{-1}$, then we expect a rate on
to a spindle pole of perhaps $5$s$^{-1}$ per microtubule.  Metazoan
cells will have large numbers of spindle microtubules in the vicinity
of an unattached kinetochore, increasing the on-rate still
further. Even if some of the signal is lost in transit to the spindle
pole, the flux is adequate to communicate the state of the
kinetochore. 
Note that the localization of the inhibitory signal means that less
amplification is needed: the flux off the kinetochore together with
directed transport are by themselves sufficient to produce good
inhibition.
The next question is how the pole processes this information. 
One attractive possibility is that the $g^*$ inhibit the spindle pole
and are subsequently released back into the cytoplasm in the inactive
$g$ form. When the last kinetochore attaches, the active transport
``wait anaphase'' signal is switched off, releasing the inhibition and
allowing the spindle pole to broadcast a ``go anaphase'' signal. This
signal could be actively transported by
plus end directed motors to communicate with connected kinetochores on
the same spindle. However, in the experiments of Ref.~\cite{rieder97},
once one spindle had entered anaphase, the other spindle also
progressed to anaphase regardless of whether it contained unattached
kinetochores. This finding suggests that a final ``go anaphase''
signal is transmitted via diffusion.

For active transport models there is little relevant experimental
data. As a result, our modelling has inevitably been more speculative
and less detailed than for diffusive sequestration models. In
particular it is not clear what the signalling molecule $g^*$ might
be.  Presumably it cannot involve BubR1-Cdc20 or Mad2-Cdc20, as we
require a short lifetime in the cytoplasm. Furthermore these molecules
are not known to bind to minus-end directed motors. The active
transport model nevertheless predicts that the inhibitory signal is
propagated away to the spindle pole by a minus-end directed motor. It
is tempting to associate the motor protein dynein with this role;
however, this assignment is problematic. If dynein were performing
this function then inhibition of dynein would effectively switch off
the inhibitory active transport ``wait anaphase'' signal, resulting in
anaphase progression.  However, experiments have revealed precisely
the opposite effect: inhibition of dynein leads to inhibition of cell
cycle progression \cite{howell01}. Furthermore this block was not due
to a more general effect of dynein inhibition, as injecting Mad2
antibodies in dynein inhibited cells still led to rapid anaphase entry
\cite{howell01}.  Hence, either dynein is not sufficiently inhibited
in this experiment, implying that some inhibitory signal can still
leak through, or else other motors are involved.  Dynein is already
known to transport kinetochore components
\cite{howell01,howell00}. However, this transport is associated with
the removal of Mad2 binding sites at kinetochores once a microtubule
has attached \cite{howell01}. 
Without these binding sites, Mad2 cannot cycle through a
kinetochore. When this removal occurs at the final attached
kinetochore, Cdc20 can no longer be sequestered by Mad2, and instead
will be free to trigger anaphase progression.

In summary, our active-transport model is consistent with
the experimental data of Rieder {\it et al.}. We can easily find a
lifetime for the inhibitory species $g^*$ that is long enough for the
$g^*$ to reach an adjacent microtubule, thereby communicating the
state of the kinetochore to the spindle pole, but short enough for
strong attenuation at another spindle $20\mu$m away.

\section*{Discussion}

\begin{figure}[t!]
\epsfig{file=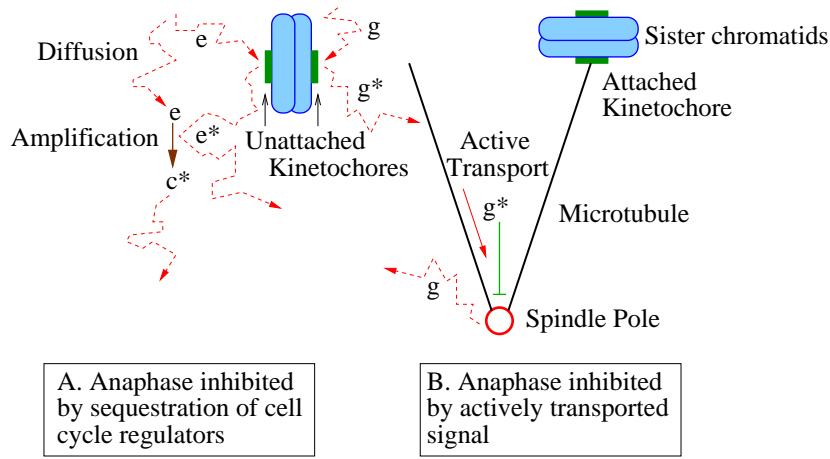,height=6.0cm}
\caption{
\lineskip 2pt \lineskiplimit 2pt
A schematic of (A) the diffusive two step reaction model and (B)
the active-transport model.  Kinetochores are shown in green, red
dashed lines denote diffusion, and solid red arrows denote motion via
active transport.  Sister chromatids are shown in blue, while the
black lines are microtubules.
\label{schem2}
}

\end{figure}

In this paper, we have shown that two models with quite different
mechanisms: the diffusive reaction cascade model, and a
model with active transport, are both possible signalling mechanisms
for the SAC. These two models are schematically illustrated in
Fig.~\ref{schem2}. 

\

\noindent
{\bf Dual Pathways for the Metazoan SAC?} One attractive possibility
to reconcile the above models and the experimental data is that, in
metazoan cells, both mechanisms are used.  Interestingly, cells with
unattached kinetochores microinjected with Mad2 antibodies prematurely
entered anaphase \cite{gorbsky98}.
This procedure will flood the cell with unsequestered Cdc20, while
unattached kinetochores will still be signalling a ``wait anaphase''
signal via the active transport mechanism. The fact that the cell
still enters anaphase
indicates that an active transport mechanism is probably not essential
for checkpoint function: the purely diffusive pathway suffices. On the
other hand, the experiments of Rieder {\it et al.} \cite{rieder97}
show that an inhibitory diffusive pathway can be
overruled by a second pathway, which, as we have seen, is likely based
on active transport. We therefore propose that {\it either} mechanism
can trigger anaphase onset, even without support from the other
pathway.
The switch off of an active transport based ``wait anaphase'' signal,
or the release of sufficient freely diffusible Cdc20, by switching off
an efficient sequestration apparatus, are each separately capable of
generating cell cycle progression.
With this assumption, our modelling is then entirely
consistent with experiment.

\

\noindent
{\bf Future work} Although some of the key principles used by the
metazoan SAC are starting to become clear, there is still much that
remains to be understood.  Even for models of signalling via
diffusion, which have been more actively pursued, further quantitative
measurements would be very useful. For example, if the {\it in vivo}
diffusion constants were found to be small then this would eliminate
some possible models.  Also, we predict that amplification without
autocatalysis is likely to be required, which implies the existence of
at least two inhibiting species ($e^*,c^*$). Of course, the {\it in
vivo} checkpoint dynamics will likely be much more complicated than
our simple outline model, but the requirement for amplification will
likely remain. Better characterization of the BubR1/Mad1/Mad2/Cdc20
protein dynamics should therefore allow these predictions to be
tested.
We would also like to emphasize the close connection between our
diffusive two step reaction and other signal cascades, such as for
MAPK.

For the alternative active transport pathway, a first goal would be to
directly observe and image its components. For example, it would
instructive to search for the transport of checkpoint proteins along
spindle microtubules {\it prior} to microtubule
attachment. Furthermore, disruption of appropriate minus end motors
may be able to generate premature anaphase onset by disrupting the
active transport inhibitory signal. As we have discussed previously,
it is also important to examine how any such signal is processed by
the spindle pole to provide inhibition.

In general, future experimental work will need to measure more of the
model parameters before we can make reliable quantitative predictions
for intracellular signalling. Nevertheless, as we have shown,
computational models can play a useful role in discriminating between
viable and inviable mechanisms of checkpoint function.

We would like to thank Fred Chang, Alexey Khodjakov, Yinghui Mao and
Kim Nasmyth for very useful discussions. MH acknowledges funding from
The Royal Society.

\end{document}